\documentclass[prl,twocolumn,showpacs,floatfix]{revtex4}
\usepackage{graphicx}
\usepackage{epstopdf}
\usepackage{color}
\usepackage{amsmath}
\usepackage{hyperref}
\usepackage{commath}

\begin{document}
\title{Pivotal role of magnetic ordering and strain in lattice thermal conductivity of chromium-trihalide monolayers}

\author{T. Pandey}
\email[]{tribhuwan.pandey@uantwerpen.be}

\author{F. M. Peeters}

\author{M. V. Milo\v{s}evi\'c}
\email[]{milorad.milosevic@uantwerpen.be}

\affiliation{Department of Physics and NANOlab Center of Excellence, University of Antwerp, Groenenborgerlaan 171, B-2020 Antwerp, Belgium}

\date{\today}

\begin{abstract}
Understanding the coupling between spin and phonons is critical for controlling the lattice thermal conductivity ($\kappa_l$) in magnetic materials, as we demonstrate here for CrX$_3$ (X = Br and I) monolayers. We show that these compounds exhibit large spin-phonon coupling (SPC), dominated by out-of-plane vibrations of Cr atoms, resulting in significantly different phonon dispersions in ferromagnetic (FM) and paramagnetic (PM) phases. Lattice thermal conductivity calculations provide additional evidence for strong SPC, where particularly large $\kappa_l$ is found for the FM phase. Most strikingly, PM and FM phases exhibit radically different behavior with tensile strain, where $\kappa_l$ increases with strain for the PM phase, and strongly decreases for the FM phase --- as we explain through analysis of phonon lifetimes and scattering rates. Taken all together, we uncover the very high significance of SPC on the phonon transport in CrX$_3$ monolayers, a result extendable to other 2D magnetic materials, that will be useful in further design of thermal spin devices.
\end{abstract}

\maketitle

\section{Introduction}
The discovery of ferromagnetic two-dimensional (2D) materials such as CrI$_3$~\cite{mcguire2015coupling,huang2017layer,lado2017origin,kashin2020orbitally,thiel2019probing}, CrBr$_3$~\cite{zhang2019direct,kim2019micromagnetometry}, CrGeTe$_3$~\cite{gong2017discovery,xu2018interplay} and other~\cite{gibertini2019magnetic,zhang2021two,och2021synthesis}, has spurred tremendous interest in controlling 2D magnetism by electric field, strain and defects. Ferromagnetism in 2D materials is highly desirable for new technological applications such as nanoscale spintronics where the electron's spin degree of freedom is used to carry information~\cite{zheng2018tunable,wei2019recent,lin2019two}. In spintronics the magnetic thermal transport can be a probe to unravel spin and topological excitations in 2D materials. Engineering thermal transport is also crucial for efficient device performance and thermal management. Particularly for magnetic devices a fundamental understanding of the spin-phonon coupling (SPC) and its effect on thermal transport is very important. 

Magnetic contributions to thermal transport are intertwined with phononic or lattice contribution and therefore the investigation of SPC is necessary to accurately determine thermal transport properties of magnetic materials. Some recent studies have explored SPC in CrX$_3$ (X= Cl, I) compounds~\cite{webster2018distinct,pocs2020giant,kozlenko2021spin} by stduing properties of paramagentic (PM), ferromagnetic (FM) and anti-ferromagnetic (AFM) phases. For magnetically ordered CrCl$_3$ it was shown experimentally that at low temperature the in-plane lattice thermal conductivity ($\kappa_l$) strongly depends on the magnetic field, a phenomenon typically known as magneto-resistance~\cite{pocs2020giant}. This giant thermal magneto-resistance is due to the suppression of spin phonon scattering under magnetic field~\cite{pocs2020giant}. Similarly, it was predicted that CrI$_3$ exhibits giant spin-lattice coupling and exhibits different lattice thermal conductivity in PM and ferromagnetic FM phases~\cite{qin2020giant}. The previous findings claim that the structural distortion caused by spin-lattice coupling are the main cause behind the high $\kappa_l$ for the FM phase. Despite these efforts the atomic origin of this strong SPC in Cr trihalide monolayers remains unknown.

Here to disentangle the SPC in Cr-trihalides we systemically study the phonon transport properties in PM, FM and AFM phases and quantify the SPC. Both CrI$_3$ and CrBr$_3$ exhibit strong SPC which we find to be mostly mediated by out of plane vibration of Cr atoms and in-plane vibrations of X atoms. Specifically, largest SPC of $\sim$ 38 cm$^{-1}$ is found for the $A_{2u}$ phonon mode. We validate the previous claim~\cite{qin2020giant} of high SPC in these compounds leading to high $\kappa_l$ for ferromagnetic phase. Additionally, by comparison of the phonon frequencies for structures with different magnetic orderings and structural distortions, we find that the \emph{SPC, is the dominant factor in the observed changes in phonon dispersion and thermal conductivity.} 

We further show that SPC can be tuned by application of tensile strain. Strikingly, depending upon the magnetic phase, strain can have diverse effect on $\kappa_l$. We find that within the strain range studied here for FM and AFM phase the $\kappa_l$ \textit{decreases} with strain, whereas for PM phase $\kappa_l$ \textit{increases} with applied strain. This observed strain-induced behavior of $\kappa_l$ is explained by calculating phonon lifetimes, and velocities. The observed anomalies in the phonon dispersions and $\kappa_l$ calculations provide clear evidence for a strong SPC in magnetic CrX$_3$ family.

\begin{figure*}[!ht]
\includegraphics[width=0.9\textwidth]{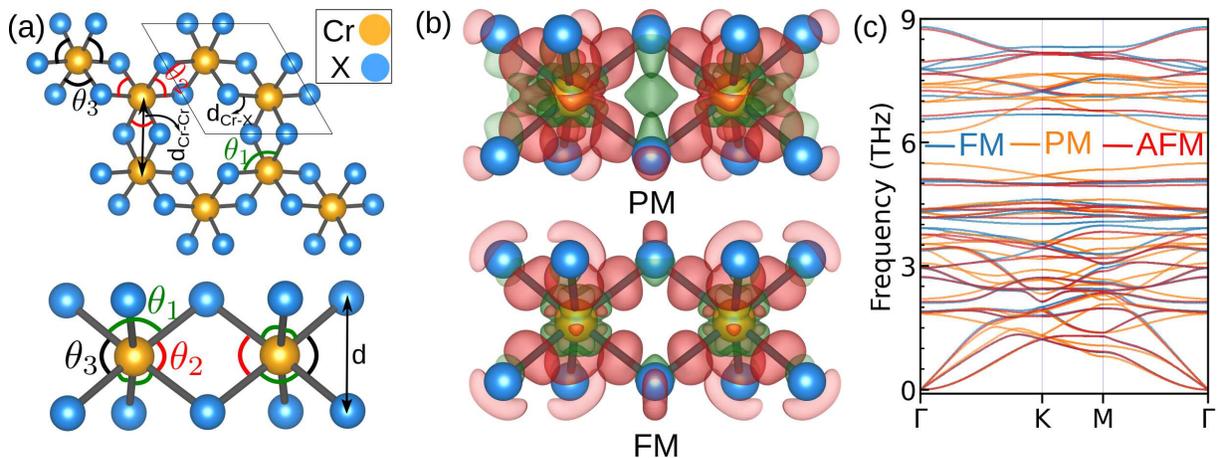}
\caption{\label{fig1} (a) Top and side view of the crystal structure of a CrX$_3$ monolayer. Different bond lengths and angles along with the hexagonal unitcell are also shown. The equivalent angles are marked by the same colors. (b) The induced charge density obtained after subtracting the atomic contributions in PM and FM phases of CrBr$_3$ plotted at an isosurface level of 0.005 eV/\AA$^3$. Green and red color represent depletion and accumulation of electrons. (c) Phonon dispersion of CrBr$_3$ monolayer in FM, AFM  and PM phases. Although in both the PM and FM (AFM) phases the acoustic phonons are comparable, significant differences can be seen at high frequencies.} 
\end{figure*}

\section{METHODOLOGY}\label{comp}
The first-principles calculations were performed based on the density functional theory (DFT) using the projector augmented wave~\cite{blochl1994projector} (PAW) method implemented in the Vienna \emph{ab-initio} simulation package (VASP)~\cite{kresse1996efficiency,kresse1999ultrasoft}. The generalized gradient approximation (GGA) from Perdew-Burke-Ernzerhof (PBE) is used for the exchange correlation functional~\cite{perdew1996generalized}. Six ($d^{5}s^{1}$) and seven ($s^{2}p^{5}$) valence electrons were used in the PAW pseudo-potential for Cr and Br/I, respectively. A large vacuum spacing of 20\AA~is employed along the out of plane direction to model monolayers. The calculations in the FM state are performed in the collinear spin configurations, whereas for the PM state spin is not considered. The spin orbit coupling is not included here.~All geometries are fully optimized until the forces on atoms are less than 0.005 eV/\AA. The energy cutoff of 500 eV, energy convergence threshold of 10$^{-6}$ eV and gamma-centered k-mesh of 15$\times$15$\times$1 was used for structural relaxation.~The relaxed lattice parameters along with bond lengths and angles are shown in Table~\ref{Table:1} for both PM and FM phase. The bi-axial tensile strain is generated by simultaneously increasing the in-plane lattice by the same ratio $\epsilon = (a - a_0)/a_0$, where \textit{a} and $a_0$ are the lattice constants of strained and unstrained structures. Subsequently all atomic positions were relaxed again. We note that under compressive strain the CrX$_3$ monolayers become dynamically unstable, therefore here we only consider tensile strain.

\begin{table*}[!ht]
\centering
\caption{The lattice constant (\textit{a}), bond lengths and bond angles for CrI$_3$ and CrBr$_3$ in PM, FM and AFM phases. These structural parameters are marked in Fig.~\ref{fig1}(a).}
\label{Table:1}
\begin{tabular}{c | c | c | c | c | c | c | c |c }
\hline
\multicolumn{1}{c|}{System} & \multicolumn{1}{c|}{Phase} & \multicolumn{1}{c|}{\textit{a} (\AA) } & \multicolumn{3}{c|}{distance (\AA)} & \multicolumn{3}{c}{angle $\angle ^{\circ}$ } \\
\cline{3-9}
{} & {} & {}& {$d_{in}$} & {$d_{Cr-X}$} & {$d_{Cr-Cr}$} & {$\theta_1$} & {$\theta_2$} & {$\theta_3$} \\[0.9ex]
 \hline
CrBr$_3$ & PM & 6.412 & 2.801 & 2.467 & 3.701 & 90.937 & 82.827 & 96.043 \\[1.5ex]
 & FM & 6.437 & 2.871 & 2.571 & 3.716 & 90.685 & 84.847 & 94.153\\[1.5ex]
 & AFM & 6.427 & 2.884 & 2.516 & 3.710 & 90.416 & 84.992 & 94.572\\[1.5ex]
\cline{1-9} 
CrI$_3$ & PM & 6.992 & 3.030 & 2.670 & 4.036 & 90.985 & 81.797 & 97.256 \\[1.5ex]
 & FM & 7.002 & 3.123 & 2.735 & 4.042 & 90.644 & 84.718 & 94.394 \\[1.5ex]
 & AFM & 6.998 & 3.1339 & 2.7335 & 4.040 & 90.406 & 84.693 &  94.950 \\[1.5ex]
\cline{1-9} 

\end{tabular}
\end{table*}

The calculation of $\kappa_l$ from first principle methods requires harmonic and anharmonic interatomic force constants (IFCs). For harmonic IFCs atoms were displaced by 0.02~\AA~using the finite displacement approach as implemented within PHONOPY~\cite{togo2015distributions,togo2015first} package. For these displaced configurations forces were calculated within DFT using the VASP code. Here, a 3$\times$3$\times$1 k-point mesh, along with energy cutoff of 500 eV and strict energy convergence criteria of 10$^{-8}$ eV were used to obtain well converged phonon frequencies. For the semiconducting phases (FM,  AFM) long-range Coulomb corrections to phonon frequencies were included using the dielectric constants and Born effective charges according to method proposed by Gonze {\it et al.}~\cite{gonze1994interatomic}. Translation and rotational invariance along with Born-Huang symmetry constraints~\cite{born1954dynamical} were imposed on the calculated harmonic IFCs following the procedure described in in Ref.~\cite{polanco2020defect}. Convergence of phonon dispersion with respect to supercell size was checke, and we find that phonon dispersions are well converged for 4$\times$4$\times$1 supercell. Therefore, for all the systems (with and without strain) the harmonic IFCs were calculated on 4$\times$4$\times$1 (128 atoms) supercell. We also checked the effect of spin orbit coupling  (SOC) on phonon dispersion for ferromagnetic phases of CrBr$_3$ and CrI$_3$. We find that SOC has no significant effect on the lattice relaxation and the phonon dispersion. Therefore, SOC is neglected for the results presented in this work. 

Anharmonic IFCs were calculated using 3$\times$3$\times$1 supercell for all systems. For the displaced supercells configurations phonon-phonon interactions were truncated at 8.5~\AA. This on average requires 500 DFT calculations for each system. To reduce the computational cost here, only $\Gamma$ point k-mesh along with energy cutoff of 450 eV and energy convergence criteria of 10$^{-7}$ eV was used for Brillouin zone integration. 

These harmonic and anharmonic force constants were used to solve the phonon Boltzmann transport equation (PBTE) iteratively. The lattice thermal conductivity ($\kappa_l$) is given by~\cite{ziman2001electrons,lindsay2019perspective}
\begin{equation}
\kappa = \kappa_{\alpha \alpha} = \frac{1}{V}\sum_{\nu} C_{\lambda}v_{\alpha,\nu}v_{\beta,\nu} \tau_{\nu},
\end{equation}
where $\nu$ defines a phonon mode with wave-vector \textbf{q} in branch $j$, and $v_{\alpha ,\nu}$ is the velocity of phonon mode $\nu$ in the Cartesian direction $\alpha$. $C_{\nu}$ is the mode specific heat and $\tau_{\nu}$ is the phonon lifetime determined from the iterative solution of PBTE. In our calculations three-phonon scattering and the isotopic disorder scattering from natural isotope mass variation are considered. The anharmonic scattering rate from the three-phonon scattering ($\tau_{\nu}^{3ph}$) processes is calculated as
\begin{equation}
\frac{1}{\tau_{\nu}^{3ph}} = \frac{1}{N}\bigg(\sum_{\nu^{\prime} \nu^{\prime\prime}}\Gamma^{+}_{\nu\nu^{\prime}\nu^{\prime\prime}} + \frac{1}{2} \sum_{\nu^{\prime} \nu^{\prime\prime}}\Gamma^{-}_{\nu\nu^{\prime}\nu^{\prime\prime}}\bigg),
\end{equation}
Here $\nu^{\prime}$ and $\nu^{\prime\prime}$ denote second and third phonon mode scattering with phonon mode $\nu$. $\Gamma^{+}_{\nu\nu^{\prime}\nu^{\prime\prime}}$ and $\Gamma^{-}_{\nu\nu^{\prime}\nu^{\prime\prime}}$ are the three-phonon scattering rates for absorption ($\nu \rightarrow \nu^{\prime} + \nu^{\prime\prime}$) and emission ($\nu + \nu^{\prime} \rightarrow \nu^{\prime\prime}$) process respectively. A Gaussian smearing width of 1.0 was used to approximate the Dirac $\delta$ distribution for the calculation of $\tau_{\nu}^{3ph}$ and a grid of 61$\times$61$\times$1 was used for the Brillouin zone integration. The isotope scattering ($\tau_{iso}$) is calculated as per Tamura's formula~\cite{tamura1983isotope, tamura1984isotope}. Finally the scattering rate of each mode is calculated by the Matthiessen's rule. These calculations were performed within the ShengBTE~\cite{ShengBTE,li2012thermal,li2012thermal-1} code. The convergence of $\kappa_l$ with respect to anharmonic force constant cutoff and integration mesh was carefully tested. We find that $\kappa_{l}$ is well converged (within 4 \%) when anharmonic IFCs are truncated at 7$^{\rm{th}}$ nearest neighbors (8.5~\AA), for a 61$\times$61$\times$1 integration mesh. $\kappa_l$ presented here is normalized with respect to the ratio of vacuum and layer thickness ($d_{in}$) listed in Table~\ref{Table:1}. The convergence test for phonon dispersion dependence on supercell size, the effect of SOC on phonon dispersions, and convergence of $\kappa_l$ with respect to integration grid, and 3rd order cutoff distance, are given in supplementary data (see Figs. S1-S3).

The electronic structure of PM and FM phases are different --- PM phase is metallic whereas FM and AFM phases are semiconducting~\cite{webster2018distinct}. This difference can be important for phonon transport, particularly in metals where electron-phonon scattering can suppress $\kappa_l$\cite{ziman2001electrons}. To test this we also calculate the $\kappa_l$ of CrBr$_3$ in the PM phase after including contributions from electron-phonon scattering. The electron-phonon scattering rates are calculated within the Electron-Phonon Wannier (EPW)~\cite{ponce2016epw} and Quantum Espresso~\cite{giannozzi2009quantum, giannozzi2020quantum} package using GBRV pseudopotential~\cite{garrity2014pseudopotentials}. Further details of these calculations are given in the supplementary data. Notably, we find that electron-phonon scattering has little effect on $\kappa_l$ of the PM phase.

\begin{figure*}[!t]
\includegraphics[width=0.85\textwidth]{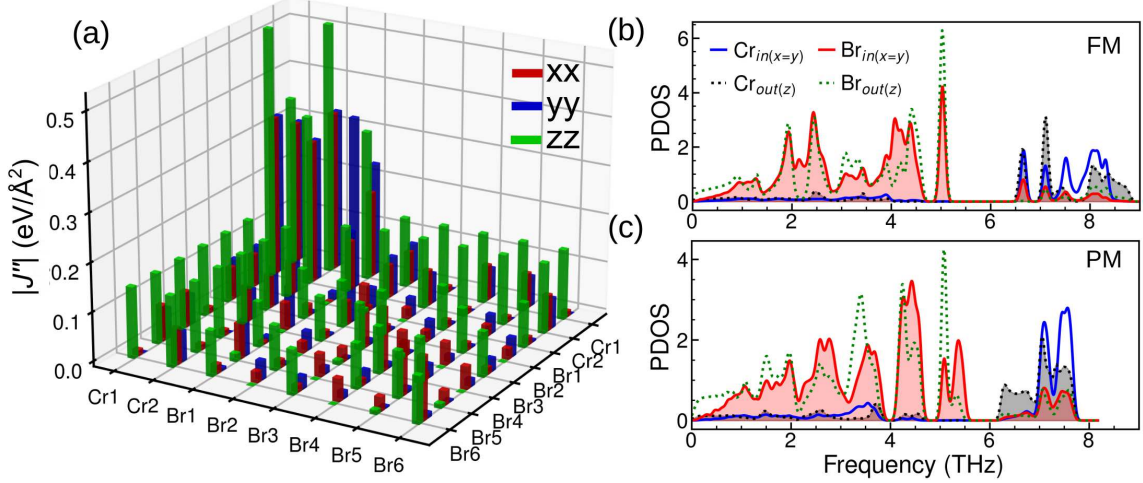}
\caption{\label{fig2} (a) Atom-resolved $J^{\prime \prime}$ contributions quantifying spin-phonon coupling of PM and FM phases in CrBr$_3$. Direction projected density of states in (b) FM and (c) PM phase. Going from PM to FM phase the in-plane vibration of Br and out-of-plane vibrations of Cr exhibit the largest change.}
\end{figure*}

\section{Effect of magnetism on structural properties and lattice dynamics}

CrX$_3$ (X = Br, I) crystallizes in the trigonal \textit{P31m} space group where the hexagonal planar lattice of Cr atoms is sandwiched between triangular planar ordering of X atoms, as shown in Fig.~\ref{fig1}(a). One Cr atom is bonded to six X atoms, forming a distorted octahedral coordination, as can be seen by different octahedral angles listed in Table~\ref{Table:1} in PM, FM and AFM phases. A relatively larger deviation from 90$^{\circ} \angle $ is seen in the PM phase octahedral angles, whereas the octahedra angles in FM and AFM phase are comparable.

The optimized lattice constants of both monolayers in the PM and FM phases are also listed in Table~\ref{Table:1}. In the FM phase we find lattice parameters of 6.437 \AA~and 7.002 \AA~for CrBr$_3$ and CrI$_3$ respectively, which match (within 1 \%) the experimentally reported lattice parameters~\cite{chen2019direct}. Lattice parameters and octahedral angles are comparable between FM and AFM phases as listed in Table~\ref{Table:1}. Depending on the presence or absence of long-range magnetism, lattice parameters also differ. For CrBr$_3$, going from PM to FM phase the lattice parameter increases by 0.42\% whereas for CrI$_3$ it increases by $\sim$ 0.15\%. Magnetism not only results in different lattice parameters but the octahedra angles are also significantly different, as shown in Table~\ref{Table:1}. Particularly the angle between top and bottom X atoms with Cr in the middle (marked as $\theta_2$ and $\theta_3$ in Fig.~\ref{fig1}(a)) is different between PM and FM phases. These structural deformations indicate the presence of spin-lattice coupling, as hinted by the previous study~\cite{qin2020giant}. 

The presence of spin order (ferro or antiferro-magnetism) also has an effect on the charge transfer between Cr and X atoms, as calculated by the Bader charge transfer~\cite{sanville2007improved,tang2009grid}. In CrBr$_3$, for the PM structure Cr donates 1.05$e^{-}$ to Br atoms, whereas in the FM (AFM) structure Cr donates higher charge of 1.25$e^{-}$ to Br atoms. This difference in charge transfer can also be seen from the induced charge density obtained after subtracting atomic contributions, shown in Fig.~\ref{fig1}(b) for both PM and FM phase of CrBr$_3$. This indicates that the bonding features in PM and FM (AFM) phases are different, which should have a direct effect on their phonon dispersion.

The calculated phonon dispersion for CrBr$_3$ in PM, FM and AFM phases are shown in Fig.~\ref{fig1}(c). As can be seen there, phonons in the FM and AFM phase are similar and exhibit comparable group velocities. When compared to the PM phase, phonons in the FM and AFM phase are stiffer and exhibit slightly larger phonon group velocity along $\Gamma$-M and $\Gamma$-K directions for all three acoustic phonon modes. In addition, for the CrBr$_3$ FM and AFM phase the gap in the optical frequency part of the spectrum (in the 4-6 THz frequency range) widens. This increased optical phonon frequency gap can impact $\kappa_l$ by reducing the phonon-phonon scattering~\cite{lindsay2015anomalous, pandey2018symmetry}. For PM-CrBr$_3$, an optical phonon branch around 6.2 THz is very dispersive and will have large group velocity. Such highly dispersive phonon branches can contribute significantly to $\kappa_l$~\cite{pandey2018symmetry}. Since the atomic masses of Cr and Br are comparable, there is large overlap in their contribution to phonon modes. The phonon dispersion and related discussion for CrI$_3$ are provided in the supplementary data (Fig. S4). Despite the structural changes listed in Table~\ref{Table:1} the phonon dispersion of PM and FM(AFM) phase in both compounds are very different. This is in contrast with the recent study on MnTe, where magnetic order induced little effect on phonon dispersions~\cite{mu2019phonons}. Previous studies on CrI$_3$ monolayers do not explain the observed differences in phonon dispersions of PM and FM phases. Therefore, to understand the interplay between magnetism and lattice dynamics we next investigate SPC in these compounds.

\section{Spin-phonon coupling}
Using the real space IFCs of the PM and FM (or AFM) phases calculated above we can extract novel insights into SPC. Within the frozen magnon method the real space IFCs for magnetic ordering can be expanded around the PM real space IFCs as~\cite{fennie2006magnetically, kumar2012spin, zhang2019first}:
\begin{equation}
\widetilde{C}_{\vec{R}{i\alpha} \vec{R^{\prime}}{j\beta}} = C_{\vec{R}{i\alpha} \vec{R^{\prime}}{j\beta}} - \sum_{mn}J^{\prime\prime}_{\vec{R}{i\alpha} \vec{R^{\prime}}{j\beta}; mn}
 \langle \vec{S}_{m} . \vec{S}_{n}\rangle,
\end{equation}
Here $\widetilde{C}$ and $C$ are the IFCs in FM and PM phases, respectively. $J^{\prime \prime}$ is the second-order derivative of exchange interaction $J$ with respect to atomic displacement. The sum runs over all magnetic atoms (Cr in this case). $\vec{R}$ and $\alpha$ refer to the displacement of the atom $i$ in the unit-cell labeled by lattice vector $\vec{R}$ along the Cartesian direction $\alpha$. Here for $J$ we only consider the first nearest-neighbor interaction as it dominates the spin dynamics in these compounds. $J^{\prime \prime}$ for FM ordering can than be obtained by $\widetilde{C}_{FM} = C - 3J^{\prime \prime}S^{2}$. Similarly for AFM phase $J^{\prime \prime}$ can be obtained as $\widetilde{C}_{AFM} = C + 3J^{\prime \prime}S^{2}$, where all three first nearest neighbors are considered and we take \textit{S} as 3/2.

The atom resolved diagonal components of $J^{\prime \prime}$ are plotted in Fig.~\ref{fig2}(a) for FM-CrBr$_3$. As can be seen from Fig.~\ref{fig2}(a), \textit{zz} component of $J^{\prime \prime}$ for Cr displacements along \textit{z} direction ($\frac{\partial J^{2}_{zz, zz}}{\partial u_{Cr1} \partial u_{Cr2}}$) is the dominating one and is over 10 times larger than any other component of $J^{\prime \prime}$. In addition, $J^{\prime \prime}$ for Br in-plane displacements ($\frac{\partial J^{2}_{xx, yy}}{\partial u_{Br1} \partial u_{Br1}}$) in also sizable. The large $J^{\prime \prime}$ value for Cr atoms indicates that the SPC coupling here is dominated by the out-of-plane Cr vibrations. This strong $J^{\prime \prime}$ \textit{zz} component for Cr displacements and \textit{xx} component for Br displacements can also explain the observed differences in the phonon dispersions between FM (AFM) and PM phases. As can be seen from Fig.~\ref{fig1}(c) the most changes in the phonon dispersion are observed in 3-5.5 THz (dominated by Br in-plane vibrations) and 6-9 THz window (dominated by Cr out-of-plane vibrations). 

To gain insights into atomic vibrations and SPC we plot the CrBr$_3$ direction projected partial density of states in Figs.~\ref{fig2}(b) and (c) for FM and PM phases, respectively. We see that within the above mentioned frequency window the in-plane (\textit{x}) vibrations of Br atoms and out-of-plane (\textit{z}) vibrations of Cr atoms are very different in PM and FM (AFM) phases, which is consistent with the behavior of $J^{\prime \prime}$. In this frequency range, due to strong SPC coupling, shifts and splittings of the FM (AFM) phonon modes from the PM modes are observed. This splitting is particularly larger for the phonon modes where Cr atoms vibrate along \textit{z} direction. The same is also true for CrI$_3$, for which the corresponding $J^{\prime \prime}$ and direction projected density of states are shown in the supplementary data (Fig. S5).

\begin{table}[!ht]
\centering
\caption{Frequencies of the $\Gamma$-point phonon vibrational modes for monolayer CrBr$_3$ in PM, FM and AFM phases. Note that frequencies of FM and AFM phases are comparable for most of the modes. The number in the parenthesis gives degeneracy of phonon mode.}
\label{Table:2}
\begin{tabular}{c | c | c | c }
\hline

 \multicolumn{1}{c|}{Mode} &  \multicolumn{3}{c}{Frequency (cm$^{-1}$) }\\

 \cline{2-4}
 \multicolumn{1}{c|}{Symmetry} &  \multicolumn{3}{c}{CrBr$_3$}  \\
\hline

 \hline
 {}     &        {PM}   &        {FM}   &        {AFM} \\[0.9ex]
 \hline
{Eg (2)}  &  62.13 &  63.37 &  62.77 \\[0.9ex]
{A2u (1)} &  73.58 &  70.46 &  71.36 \\[0.9ex]
{A1g (1)} & 102.39 &  90.89 &  91.40  \\[0.9ex]
{Eu (2)} & 113.74 &  98.31 & 98.64 \\[0.9ex]
{Eg  (2)} & 117.84 & 131.00 & 126.03 \\[0.9ex]
{A2g (1)} & 122.75 & 111.76 & 112.98 \\[0.9ex]
{Eu (2)} & 143.23 & 144.04 & 145.97 \\[0.9ex]
{Eg  (2)} & 145.73 & 139.64 & 139.30 \\[0.9ex]
{A1g (1)} &  170.24 & 169.47 & 165.86 \\[0.9ex]
{A1u} (1) & 183.16 & 168.45 & 170.63 \\[0.9ex]
{A2u} (1)& 207.87 & 293.78 & 291.92\\[0.9ex]
{Eg  (2)} & 233.11 & 259.25 & 255.34  \\[0.9ex]
{Eu (2)} & 247.52 & 235.82 & 239.69 \\[0.9ex]
{A2g} (1) & 244.97 & 219.02 & 221.61 \\[0.9ex]
\hline
\end{tabular}
\end{table}

To validate this further we focus on vibrational modes at the $\Gamma$ point in PM, FM and AFM phases. Due to the $D_{3h}$ point group symmetry of CrX$_3$ monolayer, the irreducible representation of phonon modes at the $\Gamma$ point (excluding three acoustic modes) can be written as~\cite{larson2018raman} $ \Gamma = 2A_{1g} \oplus 2A_{2g} \oplus 4E_{g} \oplus A_{1u} \oplus 2A_{2u} \oplus 3E_{u} $. The frequencies of these modes are listed in the in Table~\ref{Table:2} (See supplementary data, Table S1 for CrI$_3$). Among these 21 phonon modes, $E_g$, A$_{1g}$ and A$_{2g}$ are Raman active modes, while $E_{u}$ and $A_{2u}$ are infrared active. The rest are both Raman and infrared inactive. The atomic vibrations corresponding to the phonon modes are shown in the supplementary data (Fig. S6). As can be seen from Table~\ref{Table:2} and Table S1, between PM and FM (AFM) phases the largest splitting is observed for A$_{2u}$ and A$_{2g}$  modes. For instance the A$_{2u}$ and A$_{2g}$ phonon modes in CrBr$_3$ (CrI$_3$) exhibit a splitting of 85.9 (87.3) cm$^{-1}$ and 25.9 (32.84) cm$^{-1}$, respectively. As shown in Fig. S6 the A$_{2u}$ mode consist of in phase vibrations of Cr atoms along out of plane direction. The A$_{2g}$ mode also originates from the out-of-plane Cr vibrations, where Cr atoms vibrate with opposite phase. The observed shift of phonon modes is consistent with $J^{\prime\prime}$ calculations (Fig.~\ref{fig2}) which shows maximum contribution for Cr atoms vibrating out of plane. 

The SPC is directly proportional to the frequency shift and can be calculated using a simplified relation~\cite{sun2011magnetoelastic,lee2010strong,casto2015strong} $\omega \approx \omega_{0} + \lambda^{\prime} <S_{m} \cdot S_{n}>$. Here $\omega$ and $\omega_{0}$ are the frequency of the phonon mode and the frequency of unperturbed mode respectively. Here $\lambda^{\prime}$ is the SPC constant, with largest value of 38.1 (38.8) cm$^{-1}$ for the $A_{2u}$ mode in CrBr$_3$ (CrI$_3$) when CrX$_3$ changes from PM to FM phase. The calculated SPC in CrX$_3$ compounds is thus substantially stronger than that in other common 2D magnetic monolayers such as CrSiTe$_3$ (0.1-0.2 cm$^{-1}$)~\cite{casto2015strong}, while being comparable with NaOsO$_3$ (40 cm$^{-1}$)~\cite{calder2015enhanced} and CuO (50 cm$^{-1}$)~\cite{chen1995evidence}. As the phonons of AFM phase are comparable to the FM phase, their SPC with respect to the PM phase is also similar. If we consider frequency shift between AFM and FM phases, the largest change of phonon frequency is found for the $E_g$ mode ($\approx$ 4.7 cm$^{-1}$). This $E_g$ mode resembles shearing of the iodine atoms and the calculated frequency shift corresponds to a $\lambda$ of 2.08 cm$^{-1}$. This is comparable to the SPC of 1.2-3.19 cm$^{-1}$ calculated in Cr$_2$Ge$_2$Te$_6$ ~\cite{tian2016magneto,zhang2019first}. The above results suggest that, the SPC between the FM and the AFM phase is mainly contributed by the in-plane atomic vibrations, whereas SPC between the FM and the PM phase is dominated by the out of plane vibrations.

\begin{figure*}[!t]
\includegraphics[width=0.85\textwidth]{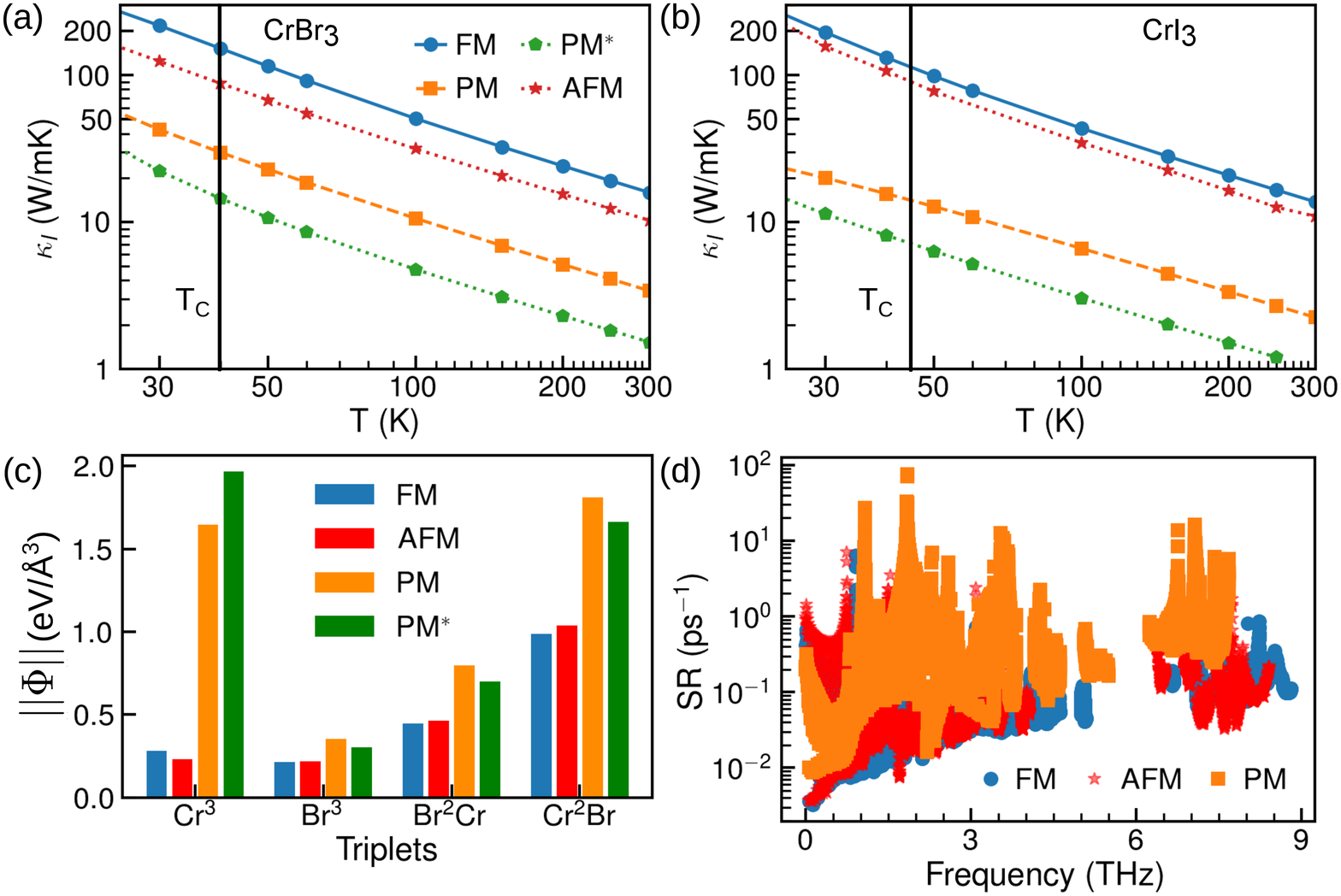}
\caption{\label{fig3} Calculated lattice thermal conductivity as a function of temperature for (a) CrBr$_3$ and (b) CrI$_3$ in both FM, PM and AFM phases. Here for comparison the $\kappa_l$ values for FM structure with no magnetic ordering (labeled as PM$^*$) are also shown for both compounds. The corresponding Curie point is marked by the black vertical line. (c) Calculated norm of the anharmonic force constants ($\norm \Phi$) for different atomic triplets in CrBr$_3$, for FM, AFM, PM and PM$^*$ phases. The IFCs are averaged over all triplets, and a large variation in anharmonic IFCs depending upon the phase considered can be seen. (d) Calculated three phonon scattering rates at 50K, for FM, AFM and PM phases of CrBr$_3$.} 
\end{figure*}

As discussed in Table~\ref{Table:1} the structural parameters within FM and PM phases are different, which could also contribute to observed differences in the phonon dispersions. To further unravel the connection between structural distortion and magnetism we consider the FM phase of CrBr$_3$ and perform phonon calculation without considering spin ordering (labeled as PM$^*$ hereafter). The resulting phonon dispersions are compared in Fig. S4, along with FM and PM phase dispersion for CrBr$_3$ and CrI$_3$. Thus ignoring the magnetic order has a significant effect on all phonon branches and their dispersions show substantial change. The maximal optical phonon frequency drops from 8.8 THz to 6.4 THz. The phonons of PM$^*$ are also quite different from the PM phase phonons. These results imply that besides the structural changes, magnetism alone can also significantly modify the phonons in these compounds. 

\section{Effect of spin order on lattice thermal conductivity}
Next, we compare the $\kappa_l$ values of CrBr$_3$ and CrI$_3$ in case of PM, FM and AFM phases, as shown in Figs.~\ref{fig3}(a) and (b), respectively. $\kappa_l$ of both compounds in the FM/AFM phase is larger than in the PM phase. For example, near the Curie point ($\sim$ 34 K) the $\kappa_l$ of CrBr$_3$ in FM phase is 155.08 W/m-K, whereas in the PM phase it is 45.56 W/m-K --- which is a difference of over three times! This means that the FM to PM phase transition will be accompanied by a striking drop in $\kappa_l$. Our results agree with the recent findings for CrI$_3$, where larger $\kappa_l$ in the FM phase was predicted~\cite{qin2020giant,liu2021effects}. To investigate the effects of spin ordering further, we also calculate $\kappa_l$ of both compounds in the PM$^*$ phase (by using the FM phase structure and neglecting the spin ordering). These calculations mimic the possibility that FM to PM phase transition does not induce any structural changes. Interestingly, as can be seen from Fig.~\ref{fig3}, neglecting magnetism has strong effect on $\kappa_l$ of the FM phase. In fact, $\kappa_l$ of PM$^*$ phase drops even below the value for the corresponding PM phase. Therefore, previous conclusions~\cite{qin2020giant} that difference in $\kappa_l$ between PM and FM phases is driven by the structural changes is only partially valid. We revealed here that \emph {magnetism alone can strongly modify the phonon spectrum and also $\kappa_l$ of these compounds via SPC}, even if one ignores the structural differences between the magnetic phases. 

To understand the variation in $\kappa_l$ depending on the magnetic phase, in Fig.~\ref{fig3}(c) we compare the norm of third-order IFCs ($\norm \Phi$) for atomic triplets in different phases. The norm is averaged over all the corresponding triplet types. These anharmonic IFCs couple with harmonic properties (phonon frequencies and eigenvectors) in the calculation of the phonon scattering rates. Since phonon scattering rates are roughly proportional to $\norm \Phi$, typically a high value of $\norm \Phi$ suggests higher anharmonicity~\cite{pandey2017lattice}. As shown in Fig.~\ref{fig3}(c), $\norm \Phi^{\mathrm{Cr^{2}Br}}$ has the maximum contribution in the FM phase. Going from FM to PM phase (or for PM$^*$ phase) the  anharmonic IFCs of Cr triplets ($\norm \Phi^{\mathrm{Cr^{3}}}$) are enhanced more than 6 times. Besides the Cr-triplet contribution, without spin ordering the anharmonic IFCs of Cr-Cr-Br triplets ($\norm \Phi^{\mathrm{Cr^{2}Br}}$) are also enhanced, implying that the PM phase is more  anharmonic than the FM phase. One key difference between FM and PM phases is that in the FM phase the strong Cr-Cr direct exchange is present due to the cation-cation coupling and the relatively lower energy of the Cr-$t_{2g}$ orbital than the Cr-$e_g$ orbital caused by the weak-field from ligand atoms~\cite{seyler2018ligand}. This results in a stronger repulsion between Cr and X atoms which makes it harder for Cr atoms to move i.e., more energy is required to move the atoms --- indicating lower anharmonicity in the FM phase. This is consistent with the computed anharmonic IFC norms discussed above. The trend of anharmonic IFC norms agrees with the calculated of anharmonic scattering rates for these phases, where higher anharmonic scattering (smaller phonon lifetime) is observed for the PM phase, as shown in Fig.~\ref{fig3}(d). We also performed $\kappa_l$ calculation in the AFM phase for both CrBr$_3$ and CrI$_3$, as shown in Figs.~\ref{fig3}(a) and (b). While in both the compounds AFM phase exhibits somewhat lower $\kappa_l$ than the FM phase, $\kappa_l$ still remains much higher compared to the PM phase. The lower $\kappa_l$ of the CrBr$_3$-AFM phase compared to the FM phase is due to the relatively larger scattering rates (smaller phonon lifetime) as shown in Fig.~\ref{fig3}(d).

Another feature of the PM phase which supports high anharmonicity is the avoided crossing between optical mode and acoustical mode along the $\Gamma$-K direction, which is absent in the FM phase. This avoided crossing is due to ratting modes and is indicative of strong coupling between optical and acoustic modes~\cite{pandey2017lattice}. The large gap at the avoided crossing point indicates a high coupling strength, which significantly increases the phonon scattering rates, ND reduces acoustic mode velocities, all of which favors a low $\kappa_l$ value. Based on the above analysis we can conclude that the low $\kappa_l$ value in the PM phase is due to (i) the large anharmonicity of Cr atoms, and (ii) avoided crossing between acoustic and optical phonons. These features mainly arise from the absence of spin order. Therefore, structural changes are not really crucial for low $\kappa_l$ of the PM phase. To the best of our knowledge, there are no lattice thermal transport measurements available to date for CrI$_3$ and CrBr$_3$ monolayers. However, our predicted trend of high $\kappa_l$ in the FM phase and low $\kappa_l$ in the PM phase agrees well with previous measurements~\cite{casto2015strong} on CrSiTe$_3$, where an increase in $\kappa_l$ was observed on cooling below the Curie temperature (T$_C$). 

\begin{figure*}[!t]
\includegraphics[width=0.97\textwidth]{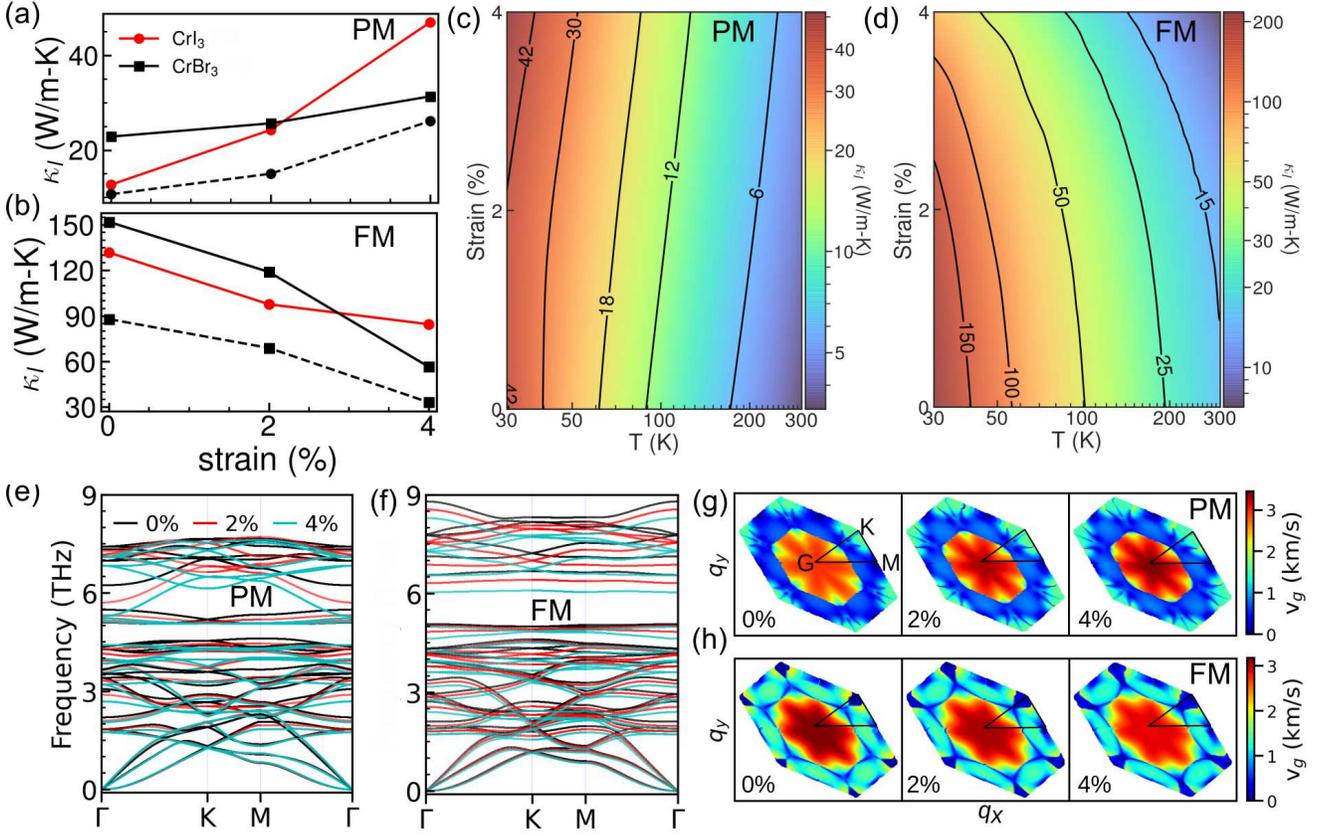}
\caption{\label{fig4}Dependence of lattice thermal conductivity ($\kappa_l$) on tensile strain, for (a) PM and (b) FM phases in CrBr$_3$ and CrI$_3$. PM and FM $\kappa_l$ is shown at 50K and 40K, respectively. Dashed lines in (a) and (b) give the strain dependent $\kappa_l$ in the PM$^*$ and AFM phase of CrBr$_3$, respectively. Contour plot of lattice thermal conductivity as a function of tensile strain and temperature is shown for CrBr$_3$ in (c) PM and (d) FM phase. Data for the contour plot is calculated at 0, 2 and 4 \% tensile strain and interpolated in between, for better visibility of the trends. At all temperatures the $\kappa_l$ values of PM and FM phases have opposite strain dependence. Tensile strain dependent phonon dispersion for (e) PM and (f) FM phases in CrBr$_3$. Variation in longitudinal acoustical phonon group velocities (${v_g}$) under tensile strain for (g) PM and (h) FM phase. In the PM phase ${v_g}$ increases with tensile strain whereas in the FM phase ${v_g}$ decreases with tensile strain.}
\end{figure*}

\begin{figure*}[!t]
\includegraphics[width=0.7\textwidth]{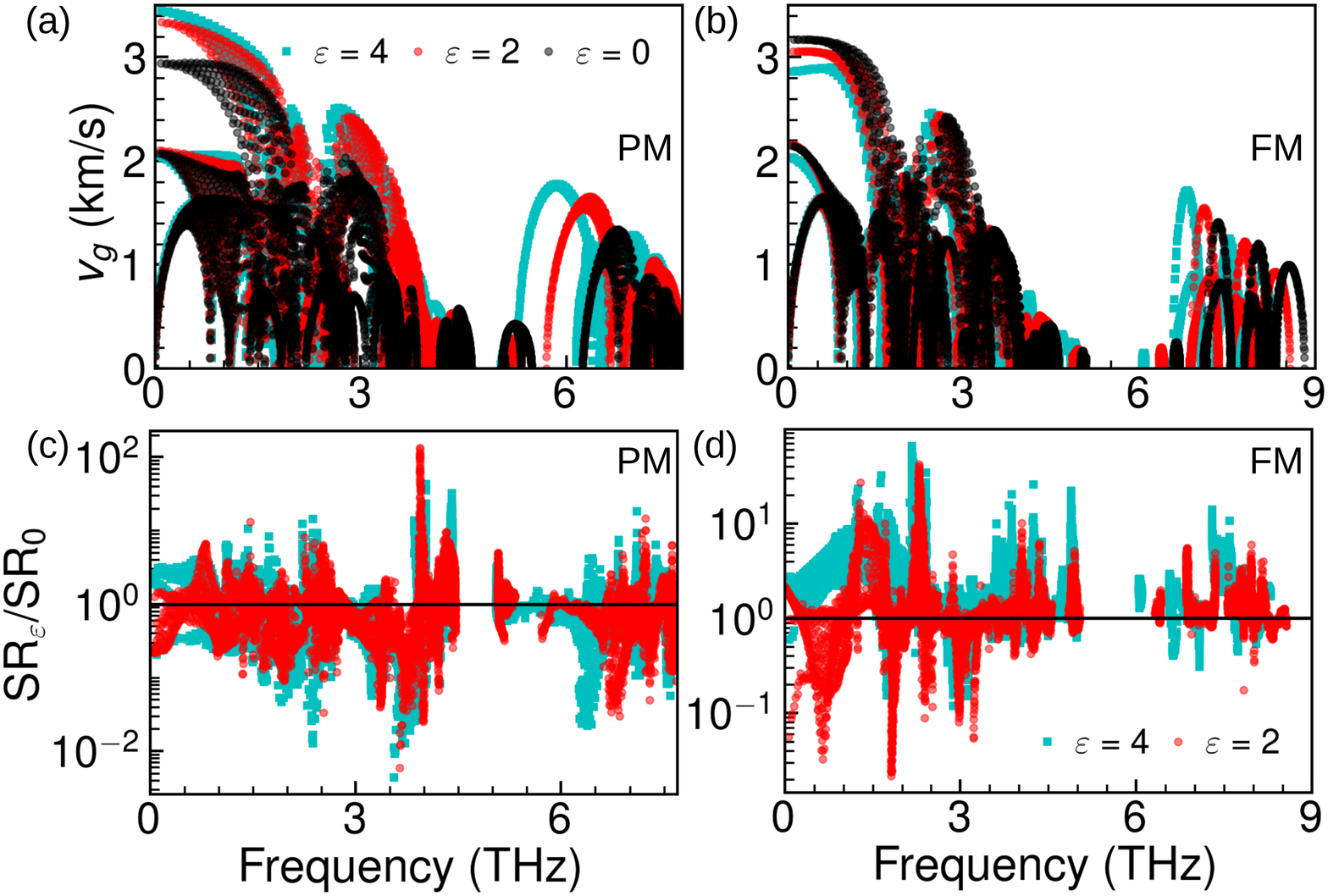}
\caption{\label{fig5} The phonon group velocity (a,b) and the three-phonon scattering rate (c,d) of PM and FM phases in CrBr$_3$, for three values of tensile strain $\varepsilon=0,2,4\%$. The scattering rates are shown normalized to the unstrained case (SR$_0$).}
\end{figure*}

\section{Effect of tensile strain on lattice thermal conductivity}

Having established that magnetic interactions or spin ordering have crucial effect on phonon transport, next we look at a possibility of tuning $\kappa_l$ by modifying the magnetic interactions. Previous reports have suggested that the magnetic properties of CrX$_3$ compounds can be tuned by an electric field~\cite{huang2018electrical,jiang2018electric,burch2018electric,xu2020electric}, magnetic field~\cite{wang2018very}, or by electrostatic doping~\cite{jiang2018controlling}. Application of strain is another convenient route to modify the magnetic exchange interaction between Cr sites in CrX$_3$ monolayers. As recently shown by Bacaksiz \textit{et al.}~\cite{bacaksiz2021distinctive}, the application of biaxial tensile strain reduces the magnetic exchange interactions and increases the Curie temperature of these compounds. Therefore, studying strain-dependent $\kappa_l$ in PM and FM ordered phases can provide further insights into the role of magnetic interactions in the behavior of $\kappa_l$, as we explore next on the example of monolayer CrX$_3$.

The dependence of $\kappa_l$ on tensile strain for CrI$_3$ and CrBr$_3$ in both PM and FM phases is shown in Figs.~\ref{fig4}(a) and (b) respectively. PM and FM $\kappa_l$ is plotted at 50K and 35K, respectively. The strain-dependent $\kappa_l$ across a wide temperature range is presented as a contour plot in Figs.~\ref{fig4}(c) and (d) for PM and FM phases of CrBr$_3$, respectively. Interestingly, depending on the magnetic ordering (PM or FM), for both compounds the strain dependence of $\kappa_l$ is reversed. Within the strain range studied here, in the PM phase $\kappa_l$ \textit{increases} with strain, whereas for the FM phase $\kappa_l$ \textit{decreases} with strain. This remains true across a wide temperature range, as seen in Figs.~\ref{fig4}(c) and (d) for CrBr$_3$. 

To analyze this distinct strain-dependent $\kappa_l$ in PM and FM phases, we next investigate the phonon dispersion and group velocity under tensile strain, which are presented in Figs.~\ref{fig4} (e)-(h). Phonon softening is observed for optical modes in both PM and FM phases. However, unlike the softening of acoustic phonons found in FM phases, in the PM phase the acoustic phonon harden. This leads to an increase in the acoustic phonon group velocity under tensile strain for the PM phase, whereas in the FM phase the phonon group velocity decreases. This is shown in Figs.~\ref{fig4}(e) and (h) for longitudinal acoustic phonons in PM and FM phases, respectively. The phonon group velocities for all phonon modes under tensile strain in both PM and FM phases of CrBr$_3$ are presented in Figs.~\ref{fig5}(a) and (b), respectively. In the PM phase within 0-5 THz frequency range the phonon group velocities are enhanced under tensile strain, whereas in the FM phase phonon group velocities decrease with increasing tensile strain. This trend of phonon group velocities agrees with the trend of $\kappa_l$ with tensile strain in PM and FM phases.

As previously stated, in the PM phase the phonon dispersions exhibit an avoided crossing between acoustic and optical phonons which results in high anharmonicity. Under tensile strain this avoided crossing is lifted, pointing to a decrease in the anharmonicity. This change in anharmonicity is further confirmed by analyzing the change in the phonon group velocities and scattering rates under tensile strain. Figs.~\ref{fig5}(c) and (d) present the three-phonon scattering rates $SR_{\varepsilon}$ under tensile strain $\varepsilon$ with respect to the scattering rate at zero strain ($SR_0$), at 50 K, in PM and FM phases, respectively. Overall in the PM phase, for the majority of phonon modes, the ratio $\frac{SR_{\varepsilon}}{SR_0}$ is lower than one, indicating a decrease in scattering rates (increase in phonon lifetime) under tensile strain. In contrast to the PM phase, in the FM phase for most of the phonon modes the ratio $\frac{SR_{\varepsilon}}{SR_0}$ is larger than one, which corresponds to an increase in the scattering rates (decrease in phonon lifetime) under tensile strain. Phonon group velocities and phonon lifetimes in the PM and FM phases of CrI$_3$ display similar behavior under tensile strain, which are shown in the supplementary data (Figs. S7-S9). The analysis presented above explains that depending up the spin order (PM or FM) the group velocity and scattering rates have different dependence on tensile strain, which translates into different $\kappa_l$ under tensile strain. To validate the role of magnetism in defining the strain dependence of $\kappa_l$ we further study the strain-dependent $\kappa_l$ in the PM$^*$ and AFM phases of CrBr$_3$, as shown in Figs.~\ref{fig4} (a) and (b), respectively. As can be seen there, similar to the PM phase, in the PM$^*$ phase $\kappa_l$ increases with strain, whereas in the AFM phase $\kappa_l$ decreases with applied strain. Further analysis shows that the strain dependence of $\kappa_l$ in the PM$^*$ and AFM phases is again dominated by trends of phonon velocities and phonon lifetimes under strain.

To gain further insights into the variation of $\kappa_l$ we look at the contribution from acoustic and optical phonon modes, shown in the supplementary data (Fig. S10). At zero strain, in both PM and FM phases acoustic phonons carry 60-80 \% of heat. Under tensile strain the contribution of acoustic phonons in the PM phase is significantly increased, mainly due to the enhanced group velocities. In the FM phase, since acoustic phonon group velocities decrease under tensile strain, their contribution to $\kappa_l$ is also suppressed. The optical phonon frequencies decrease with tensile strain and they contribute more to the scattering of phonons. This is more prominent in the FM phase, where phonon lifetimes decrease with strain. 

Next, to understand why phonons velocities and lifetimes behave differently in PM and FM phases even though both are equally strained, we looked at the variation of the internal structural parameters such as bond lengths and bond angles in both phases under tensile strain. The change in structural parameters [same as those marked in Fig.~\ref{fig1}(a)] is shown in the supplementary data (Fig. S11). Application of tensile strain decreases layer thickness $d_{in}$ in both systems due to the Poisson effect~\cite{virgin_2007}. Except for the $\theta_3$ angle, all structural parameters have similar behavior under tensile strain. However, in the PM phase under tensile strain \angle $\theta_3$ increases at a much faster rate, creating substantial deformation in the octahedral coordination.  This different behavior of the bond angle \angle $\theta_3$ contributes to observed differences in phonon dispersions under tensile strain in PM and FM phases.

\begin{figure*}[!t]
\includegraphics[width=0.9\textwidth]{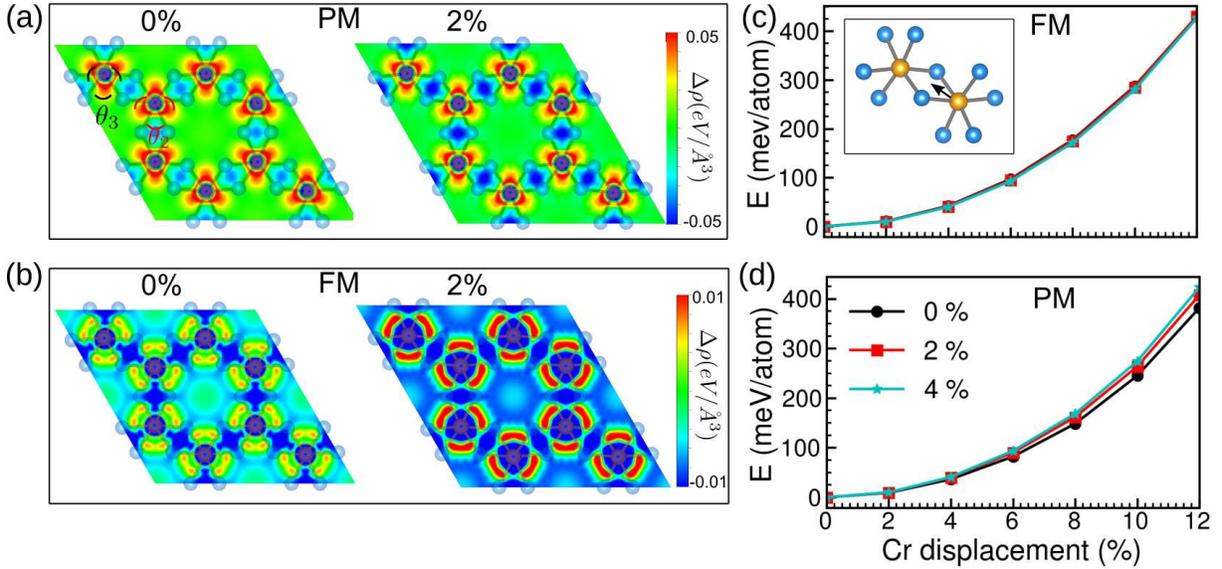}
\caption{\label{fig6} Induced charge density projected onto the (001) plane for (a) PM and (b) FM phase of  CrBr$_3$ at 0 and 2 \% strains. Energy deviation in (c) FM (d) PM phase of CrBr$_3$ when the Cr atom displaced along [010] at  some typical strains. The inset in (c) shows the direction of Cr atom displacement. }
\end{figure*}

At a first glace the strain dependence of \angle $\theta_3$ seems to be a purely structural change. There however is a direction connection with the magnetism in the system. For example in the PM$^*$ strain dependence of \angle $\theta_3$ is similar to PM phase (Fig. S11). To probe this further we next analyze the induced charge density of PM and FM phases of CrBr$_3$ under strain, which is shown in Figs.~\ref{fig6} (a) and (b), respectively. As can be seen there, the charge distribution in the FM and PM phases is very different. For the PM phase the charge accumulation around \angle $\theta_3$ is directional and localized in the central region.  In the FM phase the charge accumulation around \angle $\theta_3$ is more localized towards Cr-Br bonds. This hints that bonding is stronger in the FM phase than the PM phase, which is consistent with the higher phonon group velocity of FM phase at zero strain. When strain is applied, in the PM phase the electron accumulation around \angle $\theta_3$ does not exhibit any significant change, whereas electron depletion around \angle $\theta_2$ is increased.  Overall, the bonding strength of CrBr$_6$ octahedra is enhanced, resulting in the enhanced phonon group velocities under strain. In the FM phase, the situation is somewhat different --- significant charge redistribution is seen around the CrX$_6$ octahedra under strain. The change is particularly significant around \angle $\theta_3$. This charge redistribution in the FM phase reduces the bonding strength of CrBr$_6$ octahedra which is reflected as reduction in phonon group velocity under strain. This difference in charge transfer around \angle $\theta_3$ could also be responsible for the different strain dependence.

The different behavior of phonon lifetimes under strain in FM and PM phases can be correlated with the change in phonon anharmonicity. Phonon anharmonicity can be qualitatively characterized by the deviation of the bond energy profile from a harmonic (quadratic) behavior.  To induce the bond length change we displace one of the Cr atoms (shown in the inset of Fig.~\ref{fig6}) in both the PM and FM phases of CrBr$_3$, and compare the corresponding energy deviation for some typical strains. This is shown in the Figs.~\ref{fig6}(c) and (d). In addition to the charge redistribution seen above, in the FM phase magnetic interactions vary under strain, which can play an important role in determining the anharmonicity. When tensile strain is applied the magnetic interactions between Cr-Cr and Cr-X atoms decrease. This helps Cr atoms to move faster i.e., less energy is required to move the atoms under strain (Fig.~\ref{fig6} (c)) which leads to an increase in anharmonicity under increasing strain and eventually decreases the phonon lifetimes.  In the PM phase the movements of atoms only depend on the overlap of electronic orbitals. As discussed before, under strain both FM and PM phase behave differently, and a higher charge transfer is seen in the FM phase.  As can be seen in Fig.~\ref{fig6} (d) for PM-CrBr$_3$, the energy required to displace Cr atoms increases with strain, indicating the decrease of phonon anharmonicity and increase in phonon lifetimes. The above analysis is consistent with the change in $\norm \Phi^{\mathrm{Cr^{3}}}$ under strain. For example, the $\norm \Phi^{\mathrm{Cr^{3}}}$ in PM-CrBr$_3$ decreases from 1.81 at no strain to 1.53 eV/\AA$^3$ at 4\% strain. However, in the FM phase the $\norm \Phi^{\mathrm{Cr^{3}}}$ increases from 0.26 at no strain to 0.35 eV/\AA$^3$ at 4\% strain.

As the frequencies of phonon modes change with strain, this will also impact SPC. As described above the SPC depends on the spin-spin correlation function ($<S_{m}\cdot S_{n}>$) and shift of phonon frequencies ($\Delta \omega$). Within the strain range applied here the $<S_{m} \cdot S_{n}>$ changes  by less than 2\%, therefore the trend of SPC with strain can be described by  $\Delta \omega$. To quantify the changes in SPC under tensile strain, in the supplementary data Fig. S12 we plot the shift in phonon frequencies with respect to the PM phase under tensile strain. For the mode with largest SPC (mode $A_{2u}$) under tensile strain, the SPC is further enhanced in both CrBr$_3$ and CrI$_3$. For instance, in absence of strain the $A_{2u}$ mode in CrBr$_3$ shows splitting of 85.9 cm$^{-1}$, which under 4\% tensile strain is further enhanced to 100.5 cm$^{-1}$. This enhanced $\Delta \omega$ increases the $\lambda$ from 38.1 to 44.6 cm$^{-1}$. Additionally, several other phonon modes show large splitting between PM and FM frequencies which is further altered by strain. This frequency shift of phonon modes under strain indicates that strain can indeed be employed to tune SPC in these 2D materials.

\section{Summary}
In summary, we revealed strong effects of magnetic ordering on lattice dynamics and lattice thermal conductivity ($\kappa_l$) of monolayer CrBr$_3$ and CrI$_3$, by combining first-principle density functional theory and phonon Boltzmann transport simulations. We showed that these 2D materials exhibit large spin-phonon coupling (SPC), dominated by out-of-plane vibrations of Cr atoms, resulting in a significant change in the phonon dispersion. Specifically, a large SPC of $\sim$ 38 cm$^{-1}$ is found for the $A_{2u}$ mode. By comparing phonon dispersions for structures with different magnetic orderings and structural distortions, \textit{we find that the SPC, is a the dominant factor in the observed changes in phonon dispersions}. By analyzing the anharmonic interatomic force constants we revealed that the PM phase is more anharmonic (with smaller phonon lifetimes) than the FM phase. Together with a smaller group velocity, this high anharmonicity results in significantly smaller $\kappa_l$ in the PM phase.

Despite their very similar lattice structure, $\kappa_l$ for two magnetic orderings shows radically different response to tensile strain. For the FM phase $\kappa_l$ decreases with tensile strain, in contrast to the unusual increase observed in the PM phase. We show that the origin of this opposite trend lies in the phonon group velocities and phonon lifetimes. The response of the three-phonon scattering and phonon group velocities to tensile strain depends on the spin ordering, leading to different strain dependence of $\kappa_l$ in PM and FM phase. 

Taken together, our findings provide fundamental insights on the mechanisms behind the strong spin-phonon coupling in Cr-trihalide monolayers, and its high potential for controlling lattice thermal conduction in manners inaccessible to date. Coupled to other established manipulations of 2D materials (strain, gating, hetero-structuring), these results open a pathway to further theoretical and experimental advances regarding thermal properties of magnetic monolayers and their applications in emergent spin-caloritronic devices.

\section{Acknowledgment}
This work was supported by Research Foundation-Flanders (FWO-Vl) and the Special Research Funds of the University of Antwerp (BOF-UA). The computational resources and services used in this work were provided by the VSC (Flemish Supercomputer Center), funded by Research Foundation Flanders (FWO-Vl) and the Flemish Government department EWI. T.P. is supported by a senior postdoctoral fellowship from FWO-Vl.


\end{document}